# A safety factor approach to designing urban infrastructure for dynamic conditions

Sanjib Sharma[1], Ben Seiyon Lee[2], Robert E. Nicholas[1,3], Klaus Keller[1,4]

[1]Earth and Environmental Systems Institute, The Pennsylvania State University, University Park, PA, USA
[2]Department of Statistics, The George Mason University, Fairfax, VA, USA
[3]Department of Meteorology and Atmospheric Science, The Pennsylvania State University, University Park, PA, USA
[4]Department of Geosciences, The Pennsylvania State University, University Park, PA, USA

Corresponding author: Sanjib Sharma (sanjibsharma66@gmail.com)

**Key points:**
- We demonstrate a safety factor approach to design flood-sensitive infrastructure under changing climate and socioeconomic conditions.
- We quantify the deep uncertainty surrounding extreme rainfall projections, surface imperviousness, and infrastructure lifetime to inform a stormwater system design.
- We demonstrate how neglecting these deep uncertainties can drastically underestimate flood risks and result in poor design choices.
- We find that adding safety factors between 1.4 to 1.7 to the standard design guidance can achieve the 1/100-year hydraulic reliability for all considered cases, but at sizable additional costs.

**Abstract**

Current approaches to design flood-sensitive infrastructure typically assume a stationary rainfall distribution and neglect many uncertainties. These assumptions are inconsistent with observations that suggest intensifying extreme precipitation events and the uncertainties surrounding projections of the coupled natural-human systems. Here we demonstrate a safety factor approach to designing urban infrastructure in a changing climate. Our results show that assuming climate stationarity and neglecting deep uncertainties can drastically underestimate flood risks and lead to poor infrastructure design choices. We find that climate uncertainty dominates the socioeconomic and engineering uncertainties that impact the hydraulic reliability in stormwater drainage systems. We quantify the upfront costs needed to achieve higher hydraulic reliability and robustness against the deep uncertainties surrounding projections of rainfall, surface runoff characteristics, and infrastructure lifetime. Depending on the location, we find that adding safety factors of 1.4 to 1.7 to the standard stormwater pipe design guidance produces robust performance to the considered deep uncertainties. The insights gained from this study highlights the need for updating traditional engineering design strategies to improve infrastructure reliability under socioeconomic and environmental changes.

# 1 Introduction

Floods drive devastating climate-related impacts and human disasters (IPCC, 2021). Average global flood losses exceed US$100 billion per year (Desai et al., 2015). Model simulations project rising flood risks with intensifying climate change and increasing exposure (Hirabayashi et al., 2013; Winsemius et al., 2015). Recent catastrophic flood in Europe is a typical example how human and natural systems are exposed and vulnerable to extreme rainfall and flooding (Cornwall, 2021). A sound understanding of the transient flood risk and proper tools to account for the risk-dynamics are crucial to inform flood-risk management (IPCC, 2021; Jongman, 2018; Wong & Keller, 2017).

Engineers and decision makers around the world face nontrivial choices on how to design flood-sensitive infrastructure. Current guidelines for this decision problem (Brown et al., 2009; Schueler & Claytor, 2000) typically consider the observed historical record and assume a stationary climate (i.e., statistical properties of extremes do not change significantly over the design lifetime) (Bonnin et al., 2006; Cheng & AghaKouchak, 2015; Lopez-Cantu & Samaras, 2018; Wright et al., 2019). This stationary rainfall assumption is inconsistent with the observed record in many parts of the world (Cheng & AghaKouchak, 2015). Moreover, rainfall extremes are projected to intensify further in a warming climate (Allan & Soden, 2008). In addition, neglecting climate change violates a recent executive order in the United States (Presidential Document, 2021). This raises the question (Cheng & AghaKouchak, 2015; Wright et al., 2019; Underwood et al., 2020): How to design flood-sensitive infrastructure in a nonstationary world?

Current infrastructure design specifications typically neglect key uncertainties surrounding projections of extreme rainfall and surface runoff during infrastructure lifetime (Salas et al., 2018). Deep uncertainty surrounding rainfall projections stems, for example, from internal variability, model limitations such as unresolved processes and coarse spatio-temporal resolutions, as well as uncertainties surrounding land use change and greenhouse gas emissions scenarios (Woldemeskel et al., 2016). Deep uncertainty refers to a situation "where the system model and the input parameters to the system model are not known or widely agreed on by the stakeholders to the decision" (Lempert, 2002). In infrastructure design, for example, deep uncertainty might arise when experts and/or decision-makers cannot agree on an appropriate probability density function for describing extreme rainfall projections or a return level corresponding to a particular design return period (Lempert & Collins, 2007). Uncertainties surrounding surface runoff characteristics and the project lifetime further complicate the situation (Stehfest et al., 2019).

Previous studies (Cook et al., 2020; Lopez-Cantu & Samaras, 2018; Mallakpour et al., 2019; Wright et al., 2019) provide valuable new insights on the impacts of climate change on the design and performance of flood-sensitive infrastructures. For example, Wright et al. (2019) demonstrates that the current hydrologic design standards in the United States are insufficient due to substantial increases in extreme rainfall frequency. Mallakpour et al. (2019) shows that the hydrologic failure risk is likely to increase for most dams in California under various warming scenarios. Cook et al. (2020) analyzes how the choices in climate model spatial resolution and spatial adjustment technique alter the size of urban stormwater drainage. These studies break important new ground, but they are silent on the effects of potentially important uncertainties and their interactions (Chester et al., 2020; Manocha & Babovic, 2018; Sanders & Grant, 2020).

Here we expand on previous studies and demonstrate an approach to design stormwater infrastructure by (i) characterizing the uncertainty surrounding the project lifetime, runoff characteristics, and nonstationary extreme rainfall projections, (ii) identifying their interacting effects and relative importance on system performance, (iii) quantifying the implications of these uncertainties for the design of stormwater infrastructure, and iv) determining the safety factor required in the standard stormwater pipe design guidance to produce robust performance to the considered deep uncertainties (Figure 1). Engineers often rely on the safety-factor approach to compensate for any uncertainty in the infrastructure design process (Leitgeb, 2010; Trak, 2017). Identifying a reasonable safety factor can help to better manage the potential impacts of uncertain climatic conditions, land use characteristics and socioeconomic changes.

Specifically, we select three locations from diverse hydroclimatic regions in the United States: (i) Ellicott City (Maryland), (ii) Boulder (Colorado) and (iii) Los Angeles (California). These locations have experienced several severe flooding events over the recent decades (Smith, 2018) and represent diverse meteorological settings. Examples include floods associated with localized cloudbursts from clusters of intense thunderstorms in Ellicott City (2016 and 2018) (Viterbo et al., 2020), heavy rainfall driven floods in Boulder (2013) (Eden et al., 2016), and prolonged rainfall induced floods over California in 2017 immediately after 5 years of record-setting drought (Vahedifard et al., 2017). We use the daily rainfall records (DeGaetano et al., 2015) from gauge stations at the Baltimore/Washington International Thurgood Marshall Airport (Maryland), Boulder Municipal Airport (Colorado), and Los Angeles International Airport (California) (Supplementary Table S1).

We demonstrate the safety-factor design approach for a simple didactic decision problem: how to choose a stormwater pipe size in the face of deeply uncertain projections of extreme rainfall, surface runoff characteristics, and infrastructure lifetime (Figure 2).

## 2 Results

We first quantify the deep uncertainty surrounding extreme rainfall projections. Specifically, we estimate extreme rainfall from the historical rainfall record under the stationary assumption using a Generalized Extreme Value distribution within the Bayesian framework (see Methods Section). We find that neglecting parametric uncertainty (i.e., uncertainty associated with fitting a parametric distribution to historical data) underestimates the expected value of projected extreme rainfall events (Supplementary Figures 1b-c). Using just the best estimate of the model parameters (technically the maximum *a posteriori* probability estimate) underestimates the extreme rainfall projections by as much as 8% (Supplementary Figures 1b-c) compared to an analysis that resolves the parametric uncertainty. This effect is driven by the right-skewed return level distribution (Supplementary Figure 1b) and similar to related analyses on riverine flooding (Zarekarizi et al., 2020). The downward bias increases for longer return periods. This bias can drive higher flood infrastructure failure risks.

Accounting for nonstationarity increases the mean estimates of extreme rainfall intensity (Figures 3a, c, e). For example, the current mean estimate of the 24-hour average rainfall intensity with a 100-year return period increases by roughly 2.9 mm/hr in Boulder (Figure 3c). The increase in current mean extreme rainfall estimates increase with longer return periods. As perhaps expected, the uncertainty also increases with longer return periods. Assuming a stationary rainfall distribution for infrastructure design can lead to poor outcomes (Figures 3a, c, d). For a historical rainfall event with a stationary expected return period of 100-years, the corresponding return periods under nonstationary conditions reduces to 78 years in Ellicott City (Figure 3a), 60 years in Boulder (Figure 3c), and 90 years in Los Angeles (Figure 3e).

The extreme rainfall projections are deeply uncertain (Figures 3b, d, f). We represent this deep uncertainty by sampling from a set of (i) statistical model structures and (ii) dynamically and statistically downscaled climate models. We sample the climate models by analyzing projections from the North American Coordinated Regional Downscaling Experiment (NA-CORDEX) dataset(Mearns et al., 2017) and the Multivariate Adaptive Constructed Analogs (MACAv2-METDATA) dataset (Abatzoglou & Brown, 2012). Our projections sample different climate model structures, model resolutions, and downscaling methods. These differences lead to sizeable variation in the extreme rainfall projections (Figures 3b, d, f). Choosing just a subset of the considered climate projections can drastically undersample the apparent deep uncertainties.

The reliability of stormwater infrastructure designs can be highly sensitive to the considered deep uncertainties (Figure 4). We quantify reliability for this stormwater drainage pipe design as the probability that it does not flood over the specified lifetime (Tran Huu D., 2016). The projected reliabilities are deeply uncertain, as they are impacted by deep uncertainty projections of rainfall, surface runoff characteristics and infrastructure lifetime. The key driver of this uncertainty surrounding hydraulic reliability is the deep uncertainty surrounding extreme rainfall projections (Supplemental Table S4). Designing infrastructure in the face of deep and dynamic uncertainties poses, of course, highly complex decision problems (Bryant & Lempert, 2010; Chester et al., 2020; Herman et al., 2015; Lopez-Cantu & Samaras, 2018). Using the simple and salient example of designing a stormwater drainage pipe, we demonstrate that climate uncertainty is the dominant driver of the decision-relevant metric of hydraulic reliability.

We show how increasing investments can increase reliability and robustness against deep uncertainty. In our simple examples, adding safety factors of 1.4 to 1.7 to the pipe diameter from the standard design guidance roughly achieves the 1/100-year hydraulic reliability in a way that is robust against the considered deep uncertainties in all three considered locations (Figure 4). Safety factors of 2.0 to 2.3 can achieve the 1/500-year hydraulic reliability (Figure 4). This is, of course, just a hypothetical case for selected locations in the U.S.A. Expanding this simple analysis to inform the design of workable and acceptable design guidance poses nontrivial research questions at the interface between disciplines such as Earth sciences, engineering, economics, ethics, political science, and law. The proposed safety factor approach can potentially inform practitioners and decision makers to update design standards for keeping infrastructure reliable under dynamic climate, land use and socioeconomic uncertainties.

# 3 METHODS

## 3.1 Dataset description

We use the historical rainfall and temperature station observation from the United States National Oceanic and Atmospheric Administration (NOAA) National Center for Environmental Information data portal (https://www.ncdc.noaa.gov/ ).

We consider potential physical drivers of extreme rainfall as covariates for describing changes in statistics of extremes. Specifically, we use the historical winter mean (JFM) North Atlantic oscillation (NAO) index (Jones et al., 1997) as a covariate for the nonstationary GEV model described below. For the Atlantic hurricane covariate time series, we use the number of Atlantic tropical cyclones from the United States National Oceanic and Atmospheric Administration (NOAA) Atlantic Oceanographic and Meteorological Laboratory data portal (https://www.aoml.noaa.gov/hrd/tcfaq/E11.html ).

For the historical temperature time series, we use the daily air temperature record (DeGaetano et al., 2015) from a station at the Baltimore/Washington International Thurgood Marshall Airport (BWI), Maryland, USA. We adopt the annual global mean surface air temperature dataset from the NOAA National Centers for Environmental Information data portal (https://www.ncdc.noaa.gov/ ). We use the Atlantic main development region (MDR) sea surface temperature from the National Centers for Environmental Information data portal (https://www.ncdc.noaa.gov/), and as projection we use the coupled model intercomparison project phase 5 (CMIP5) Global climate model (GCM) (Taylor et al., 2012), namely, GFDL-ESM2M under Representative Concentration Pathway 8.5 (RCP8.5). We caution that our analysis does not account for internal variability, as well as model structural or parametric uncertainty regarding future projections of selected covariates. Assessing the impacts of these uncertainties on projected extreme rainfall is another important avenue to improve this current approach.

We use sea surface temperature anomalies in the NINO3.4 region (averaged over $5^0$N-$5^0$S, $120^0$W-$170^0$W) (Reynolds et al., 2002) available on the NOAA National Centers for Environmental Information data portal (https://www.ncdc.noaa.gov/teleconnections/enso/indicators/sst/ ), and as projection we use the coupled model intercomparison project phase 5 (CMIP5) Global climate model (GCM), namely, GFDL-ESM2M under Representative Concentration Pathway 8.5 (RCP8.5). This region provides a good measure of important changes in sea surface temperature and its gradients that result in changes in the pattern of deep tropical convection and atmospheric circulation.

We use the Pacific decadal oscillation (PDO) index (Mantua et al., 1997) and Southern oscillation index (SOI) (Ropelewski & Jones, 1987) available on the NOAA National Weather Service Climate Prediction Center data portal (https://www.cpc.ncep.noaa.gov/data/teledoc/telecontents.shtml ). The SOI represents the large-scale fluctuations in air pressure between the western and eastern tropical Pacific.

We incorporate rainfall time series from the Multivariate Adaptive Constructed Analogs (MACA) dataset (Abatzoglou & Brown, 2012). The MACAv2-METDATA dataset is the statistically downscaled subset of the coupled model intercomparison project phase 5 (CMIP5) Global climate models (GCMs). The MACA outputs include CCSM4, CanESM2, GFDL-ESM2M, inmcm4, MIROC5, and MIROC-ESM.

We adopt rainfall time series from the North American Coordinated Regional Downscaling Experiment (NA-CORDEX) project available at 25-km resolution (Mearns et al., 2017). The NA-CORDEX dataset is the dynamically downscaled dataset. The NA-CORDEX dataset corresponds to a collection of high-resolution projections from different GCM-Regional climate model (RCM) combinations. The NA-CORDEX outputs include RegCM4 driven by MPI-ESM-LR, WRF driven by MPI-ESM-LR, and WRF driven by GFDL-ESM2M under Representative Concentration Pathway 8.5 (RCP8.5). We employ the non-parametric quantile mapping approach, called Kernel

Density Distribution Mapping (McGinnis et al., 2015), to bias-correct the time series to match observations (Cook et al., 2020). We select these datasets for this proof-of-concept because they all are publicly available, frequently used for climate change impact assessments, and often used to inform flood-sensitive infrastructure design (Cook et al., 2020; Lopez-Cantu et al., 2020; Mallakpour et al., 2019). We caution that our analysis relies on selected GCMs and thus likely undersamples the full range of relevant uncertainties. Considering a richer sample of this specific structural uncertainty is a way to refine this study.

**3.2 Extreme value model**

We model the extreme rainfall using a nonstationary Generalized Extreme Value (GEV) distribution. The GEV permits accounting for the potential nonstationarity in annual flood peaks by specifying time-dependent model parameters. The GEV distribution includes the location ($\mu$), scale ($\sigma$), and shape parameter ($\xi$) to specify the center, spread, and tail behavior, respectively. The probability density function of the GEV distribution is:

$$f(x) = \begin{bmatrix} \frac{1}{\sigma}\left(1 + \xi\frac{x-\mu}{\sigma}\right)^{-1-\frac{1}{\xi}} e^{-\left(1+\xi\frac{x-\mu}{\sigma}\right)^{-\frac{1}{\xi}}}, & for \quad 1 + \xi\frac{x-\mu}{\sigma} > 0 \\ \frac{1}{\sigma} e^{\frac{\mu-x}{\sigma} - e^{\frac{\mu-x}{\sigma}}}, & for \quad \xi = 0 \\ 0, & otherwise \end{bmatrix} \quad (1)$$

We can incorporate both heavy and light tails in the GEV distribution. Based on the shape parameter, the GEV can take one of three forms: i) Frechet, or lower bounded with a heavy upper-tailed, if $\xi$ is positive; ii) Gumbel, or unbounded with heavy upper tails, if $\xi$ is zero; iii) "Reverse" Weibull, or upper bounded with heavy lower tails, if $\xi$ is negative (Coles, 2001).

We incorporate potential nonstationarity into the GEV model by allowing the location parameter ($\mu$) to covary with different physical processes (*T*) (Ragno et al., 2019):

$$\mu = \mu_0 (1 + a_\mu T), \tag{2}$$

where $T$ is the covariate, $\mu_0$ is the location parameter when $T=0$, and $a_\mu$ is the sensitivity of location parameter with respect to changes in the covariate.

We use a Bayesian approach to fit the nonstationary GEV distributions to the annual maximum rainfall intensity. Bayes' theorem combines the prior information on model parameters and likelihood function into the posterior distribution of the parameters, given the data $(p(\theta|x))$:

$$p(\theta|x) \propto L(\theta)p(\theta), \tag{3}$$

where $L(\theta)$ is the likelihood function and $p(\theta)$ is the prior distribution of random variable $\theta$. Our objective is to approximate the posterior distribution $(p(\theta|x))$ by drawing samples via Markov chain Monte Carlo (MCMC). For a random $X = \{x_1, x_2, \ldots, x_n\}$, the likelihood function $L(\theta)$ for the parameter vector $\theta$ associated with its PDF $f_x(x)$ is defined as:

$$L(\theta) = \prod_{i=1}^{n} f_x(x_i|\theta). \tag{4}$$

We use a Gaussian prior distribution centered at 0, with a wide variance $N(0,100)$ for each model parameter. We sample from the posterior distribution $(p(\theta|x))$ of the model parameters using the Metropolis-Hastings algorithm (Chib & Greenberg, 1995). We sample each GEV parameter successively for 50,000 iterations. The first 10,000 iterations are discarded for burn-in. We use the remaining 40,000 samples to serve as the ensemble for analysis. The best guess estimate refers to the parameter set with the highest posterior density value among all MCMC samples. To account for the uncertainty of rainfall intensity, we consider the full ensemble of parameter samples.

We construct a nonstationary GEV model that allows integrating relevant physical drivers of extreme rainfall as covariates. Incorporating physical-related covariates into the GEV distribution assists the statistical model to avoid unrealistic extrapolations.

### 3.3 Goodness-of-fit measures

We select several metrics for model comparison. We compute two penalized-likelihood criteria- the Akaike information criterion (Akaike, 1974) (AIC) and the Deviance information criterion (Spiegelhalter et al., 2002) (DIC). The AIC and DIC penalize models based on their goodness-of-fit as well as the effective number of model parameters. For all three criteria, models with smaller values are typically preferred over those with larger values. The AIC is defined as

$$AIC = -2\log(Lmax) + 2N_p, \tag{5}$$

where, $Lmax$ denotes the maximum value of the likelihood function within the posterior model ensemble and $N_p$ is the number of model parameters.

The deviance for a given set of model parameters is given by

$$D(\theta) = -2\log(L(\theta)). \tag{6}$$

We denote the expected value of $D(\theta)$ over $\theta$ as $\bar{D}$, and $\bar{\theta}$ as the expected value of $\theta$. We calculate the effective number of parameters as $P_D = \bar{D} - D(\bar{\theta})$. The DIC is then defined as

$$DIC = P_D + \bar{D}. \tag{7}$$

### 3.4 Covariate selection

Previous studies (Jones et al., 1997; Lapp et al., 2012; Ropelewski & Jones, 1987) show that the sea surface temperature variability has a strong influence on the variability of atmospheric climate through atmospheric teleconnections. Following previous research (Ragno et al., 2019; Wong et al., 2018; Wong & Keller, 2017), we construct a nonstationary statistical model that integrates relevant physical drivers of extreme rainfall as covariates (Figure 3): global mean surface temperature, local surface temperature, Atlantic main development region temperature (Grinsted et al., 2013), North Atlantic Oscillation index (Jones et al., 1997), Pacific Decadal Oscillation index (Mantua et al., 1997), Nino 3.4 index (Reynolds et al., 2002), and Southern Oscillation index (Ropelewski & Jones, 1987). We use the Akaike information (Akaike, 1974) and Deviance information (Spiegelhalter et al., 2002) criteria (AIC and DIC) to identify the best model structure and covariate that fits the observed rainfall dataset. Our analysis suggests that the extreme rainfall distribution in Ellicott City reflects a greater contribution from the average sea surface temperature in the Atlantic main development region (Grinsted et al., 2013) (Supplementary Table S2). The Atlantic main development region ($10^0$N-$20^0$N, $80^0$W-$20^0$W) is monitored for potential tropical system development regions during the course of a given Atlantic hurricane season. For the observed rainfall distribution in Boulder, local temperature emerges as the best covariate choice. We estimate extreme rainfall intensity in Los Angeles by taking advantage of the dependence relationship in the time series of the Nino 3.4 index and historical rainfall observations. Previous studies (Hoell et al., 2016; Wise et al., 2015) suggest that the Nino 3.4 average sea surface temperature anomalies in the area of tropical Pacific Ocean (5°S–5°N, 120°–170°W) has a substantial influence on precipitation in many parts of California.

### 3.5 Reliability assessment

We assess the reliability of hydraulic performance of stormwater drainage pipe. From reliability theory (Melchers & Beck, 2018), a limit state function ($G$) is defined as

$$G(q) = Q_p - Q_y, \quad (8)$$

where, $q$ is the random variable that represents, for example, the pipe's flow in our study, $Q_p$ is the pipe's flow capacity and $Q_y$ is the flow load. The failure probability ($P_f$) can be expressed as

$$P_f = probability\ [G(q) < 0]. \quad (9)$$

Hydraulic reliability ($R$) of a stormwater pipe design is defined as

$$R = 1 - P_f. \quad (10)$$

For a hydraulic design of stormwater pipes, we assume that the peak flow should not exceed the flow capacity of pipes. We use the peak flow from the rational method (Mays, 2010) to estimate flow load:

$$Q_y = 0.278CIA, \qquad (11)$$

where, $Q_y$ is the flow load (m³/s), $C$ is the runoff coefficient (dimensionless), $I$ is rainfall intensity (mm/hr), and $A$ is the contributing drainage area (km²). The runoff coefficient represents the integrated effects of infiltration, land use land cover, ground slopes, and soil types. It indicates the amount of runoff generated given an average intensity of rainfall for a storm.

We design the stormwater pipes for open-channel flow. We use the Manning's equation[53] to estimate the flow capacity of a circular pipe running full, but not under pressure:

$$Q_p = \frac{0.31}{n} D^{8/3} S^{1/2}, \qquad (12)$$

where, $n$ is the Manning's roughness coefficient representing the resistance to flows in pipe, $D$ is the pipe diameter (m), and $S$ is the pipe bottom slope. We consider a concrete pipe with Manning's roughness, $n = 0.013$, and a contributing drainage area of 0.25 km².

### 3.6 Uncertainty decomposition

We employ the cumulative uncertainty approach (Kim et al., 2019) to assess the influential uncertainty sources in the hydraulic reliability of stormwater pipe design. The approach decomposes the total uncertainty to individual uncertainty sources, such that the sum of the uncertainties from individual sources is always equal to the total uncertainty in the reliability of the pipe. We consider uncertainty from three key sources: climate, surface runoff characteristics, and service life of the pipe. We quantify both the individual and combined sources of uncertainty. We consider nine sets of climate model outputs, four sets of surface runoff characteristics and three sets of pipe service life (Supplemental Table S3). This produces a total of 108 scenarios.

We assess the uncertainty contribution from each source (climate, surface runoff characteristics, and service life of the pipe), called here as stage uncertainty. Stage uncertainty is the sum of the variation of the main effect of stage $k$ and the variations of the interactions between stage $k$ and stages after $k$. To quantify the uncertainty contribution from each stage, we first compute the conditional cumulative uncertainty up to a particular stage. Conditional cumulative uncertainty up to a particular stage represents the variation in the reliability due to the design choices up to that stage, while the choices beyond that stage are fixed. Then the marginal cumulative uncertainty up to a particular stage is an average of conditional cumulative uncertainties. We compute the uncertainty of each stage as the difference between successive marginal cumulative uncertainties.

We denote the total number of stages in the reliability estimates by $K$, where in this case $K=3$, i.e., climate, surface runoff characteristics, and service life of the pipe. For a particular stage $k$, there are $n_k$ models/scenarios denoted by $\chi_k$. The cumulative uncertainty up to stage $k$ is defined as the variation in the reliability due to the choice of models/scenarios up to stage $k$, while the models/scenarios after stage $k$ are fixed. The cumulative uncertainty up to stage $k$ is denoted by $U^{cum}(\chi_1,...,\chi_k)$. For a specific model/scenario, of stage $k$ for $k=1,...,K$, we let $P(x_1, x_2,...,x_K)$ be the reliability of stormwater drainage design using the models/scenarios $x_k, k = 1,...,K$. For a given model/scenario, after stage $k$, the set of reliabilities are:

$$q_{x_{k+1},...,x_K} = \{P(x_1,...,x_k, x_{k+1},...,x_K): x_j \in \chi_j, j = 1,...,k\}. \quad (13)$$

Then $U^{cum}(q_{x_{k+1},...,x_K})$ is the conditional cumulative uncertainty up to stage $k$ while the models/scenarios after stage $k$ are fixed as $x_{k+1},...,x_K$. We define the marginal cumulative uncertainty up to stage $k$ as the average of conditional cumulative uncertainties:

$$U^{cum}(\chi_1,...,\chi_k) = \frac{1}{\prod_{j=k+1}^{K} n_j} \sum_{x_{k+1} \in \chi_{k+1}} \cdots \sum_{x_K \in \chi_K} U(q_{x_{k+1},...,x_k}). \quad (14)$$

Since the cumulative uncertainty is monotonously increasing (Kim et al., 2019), we can define the uncertainty of each stage as the difference between successive cumulative uncertainties. That is, the uncertainty of stage $k$, denoted by $U^{cum}(\chi_k)$, can be defined as:

$$U^{cum}(\chi_k) = U^{cum}(\chi_1, \ldots, \chi_k) - U^{cum}(\chi_1, \ldots, \chi_{k-1}). \qquad (15)$$

We express both the stage and cumulative uncertainties in terms of the variance (Bosshard et al., 2013) in the reliability of stormwater pipe design. For $y = q_{x_{k+1},\ldots,x_K}$ and a set of $y = \{y_1, \ldots, y_n\}$, the variance is defined as:

$$\text{Variance} = \frac{1}{n} \sum_{i=1}^{n} (y_i - \bar{y})^2, \qquad (16)$$

where, $\bar{y} = \frac{1}{n} \sum_{i=1}^{n} y_i$.

The cumulative uncertainty approach has several advantages over the commonly used analysis of variance (ANOVA) framework. The ANOVA approach partitions the total variance into to the individual sources and their interaction (Bosshard et al., 2013). The ANOVA approach has several key limitations (Kim et al., 2019): (1) sensitive to outliers; (2) assumes the projections are normally distributed; (3) model selection is challenging; and (4) difficult to characterize how interaction effects drive uncertainty in projections. The cumulative uncertainty approach allows us to quantify the relative contribution of each stage to the total uncertainty in the reliability estimates and to identify how uncertainties are propagated as the stages proceed in the stormwater pipe design.

**Acknowledgements**

This work was co-supported by the National Oceanic and Atmospheric Administration (NOAA) through the Mid-Atlantic Regional Integrated Sciences and Assessments (MARISA) program under NOAA grant NA16OAR4310179 and by the Penn State Center for Climate Risk Management. All errors and opinions are from the authors and do not reflect the funding agencies. We thank Amir AghaKouchak for helpful comments on a previous draft. We thank Lauren Cook, Costantine

Samaras, and Seth McGinnis for sharing the North American Coordinated Regional Downscaling Experiment (NA-CORDEX) dataset. We are grateful to Irene Schaperdoth, Tony Wong, Murali Haran, and Skip Wishbone for valuable inputs.

**Disclaimer and License**

**Author contributions**

All authors contributed to the study design. S.S. and B.L. constructed the nonstationary statistical model. S.S. led the calculations. S.S. and K.K wrote the initial draft of the manuscript. All authors revised and edited the manuscript.

**Data and Code Availability**

The code used for this analysis and the data required to plot the results is available at GitHub repository https://github.com/svs6308/PipeDesign. All data and code currently available at GitHub will be published via Zenodo upon article acceptance.

**Competing interests**

The authors are not aware of any competing financial or nonfinancial interests.

**Materials & Correspondence**

Correspondence and requests for materials should be addressed to the corresponding author at sanjibsharma66@gmail.com

**References**

Abatzoglou, J. T., & Brown, T. J. (2012). A comparison of statistical downscaling methods suited for wildfire applications. *International Journal of Climatology*, *32*(5), 772–780.

Akaike, H. (1974). A new look at the statistical model identification. *IEEE Transactions on Automatic Control*, *19*(6), 716–723.

Allan, R. P., & Soden, B. J. (2008). Atmospheric warming and the amplification of precipitation extremes. *Science*, *321*(5895), 1481–1484.

Bonnin, G. M., Martin, D., Lin, B., Parzybok, T., Yekta, M., & Riley, D. (2006). Precipitation-frequency atlas of the United States: NOAA Atlas 14, volume 1, version 4. *NOAA, National Weather Service, Silver Spring, Maryland*.

Bosshard, T., Carambia, M., Goergen, K., Kotlarski, S., Krahe, P., Zappa, M., & Schär, C. (2013). Quantifying uncertainty sources in an ensemble of hydrological climate-impact projections. *Water Resources Research*, *49*(3), 1523–1536.

Brown, S. A., Schall, J. D., Morris, J. L., Stein, S., Warner, J. C., & Others. (2009). *Urban drainage design manual: hydraulic engineering circular 22*. National Highway Institute (US). https://rosap.ntl.bts.gov/view/dot/44353

Bryant, B. P., & Lempert, R. J. (2010). Thinking inside the box: A participatory, computer-assisted approach to scenario discovery. *Technological Forecasting and Social Change*, *77*(1), 34–49.

Cheng, L., & AghaKouchak, A. (2015). Nonstationary Precipitation Intensity-Duration-Frequency Curves for Infrastructure Design in a Changing Climate. In *Scientific Reports* (Vol. 4, Issue 1). https://doi.org/10.1038/srep07093

Chester, M. V., Shane Underwood, B., & Samaras, C. (2020). Keeping infrastructure reliable under climate

uncertainty. In *Nature Climate Change*. https://doi.org/10.1038/s41558-020-0741-0

Chib, S., & Greenberg, E. (1995). Understanding the Metropolis-Hastings Algorithm. In *The American Statistician* (Vol. 49, Issue 4, pp. 327–335). https://doi.org/10.1080/00031305.1995.10476177

Coles, S. (2001). *An Introduction to Statistical Modeling of Extreme Values*. Springer, London.

Cornwall, W. (2021). Europe's deadly floods leave scientist stunned. In *Science*, (Vol. 373, Issue 6553, pp. 372-373). 10.1126/science.373.6553.372

Cook, L. M., McGinnis, S., & Samaras, C. (2020). The effect of modeling choices on updating intensity-duration-frequency curves and stormwater infrastructure designs for climate change. *Climatic Change*, *159*(2), 289–308.

DeGaetano, A. T., Noon, W., & Eggleston, K. L. (2015). Efficient Access to Climate Products using ACIS Web Services. In *Bulletin of the American Meteorological Society* (Vol. 96, Issue 2, pp. 173–180). https://doi.org/10.1175/bams-d-13-00032.1

Desai, B., Maskrey, A., Peduzzi, P., De Bono, A., & Herold, C. (2015). *Making development sustainable: the future of disaster risk management, global assessment report on disaster risk reduction*. https://archive-ouverte.unige.ch/unige:78299

Eden, J. M., Wolter, K., & Otto, F. E. L. (2016). Multi-method attribution analysis of extreme precipitation in Boulder, Colorado. *The Environmentalist*. https://iopscience.iop.org/article/10.1088/1748-9326/11/12/124009/meta

Grinsted, A., Moore, J. C., & Jevrejeva, S. (2013). Projected Atlantic hurricane surge threat from rising temperatures. *Proceedings of the National Academy of Sciences of the United States of America*, *110*(14), 5369–5373.

Heaney, J. P., Sample, D., & Wright, L. (2002). *Costs of Urban Stormwater Control*. U.S. Environmental Protection Agency, Office of Research and Development, National Risk Management Research Laboratory.

Herman, J. D., Reed, P. M., Zeff, H. B., & Characklis, G. W. (2015). How should robustness be defined for water systems planning under change? *Journal of Water Resources Planning and Management*,

*141*(10), 04015012.

Hirabayashi, Y., Mahendran, R., Koirala, S., Konoshima, L., Yamazaki, D., Watanabe, S., Kim, H., & Kanae, S. (2013). Global flood risk under climate change. *Nature Climate Change*, *3*(9), 816–821.

Hoell, A., Hoerling, M., Eischeid, J., Wolter, K., Dole, R., Perlwitz, J., Xu, T., & Cheng, L. (2016). Does El Niño intensity matter for California precipitation? In *Geophysical Research Letters* (Vol. 43, Issue 2, pp. 819–825). https://doi.org/10.1002/2015gl067102

IPCC (2021). Climate Change 2021: *The Physical Science Basis. Contribution of Working Group I to the Sixth Assessment Report of the Intergovernmental Panel on Climate Change* [Masson-Delmotte, V., P. Zhai, A. Pirani, S. L. Connors, C. Péan, S. Berger, N. Caud, Y. Chen, L. Goldfarb, M. I. Gomis, M. Huang, K. Leitzell, E. Lonnoy, J. B. R. Matthews, T. K. Maycock, T. Waterfield, O. Yelekçi, R. Yu and B. Zhou (eds.)]. Cambridge University Press.

Jones, P. D., Jónsson, T., & Wheeler, D. (1997). Extension to the North Atlantic Oscillation using early instrumental pressure observations from Gibraltar and south-west Iceland. *International Journal of Climatology: A Journal of the Royal Meteorological Society*, *17*(13), 1433–1450.

Jongman, B. (2018). Effective adaptation to rising flood risk. *Nature Communications*, *9*(1), 1986.

Kass, R. E., & Raftery, A. E. (1995). Bayes Factors. *Journal of the American Statistical Association*, *90*(430), 773–795.

Kim, Y., Ohn, I., Lee, J.-K., & Kim, Y.-O. (2019). Generalizing uncertainty decomposition theory in climate change impact assessments. *Journal of Hydrology X*, *3*, 100024.

Lapp, S. L., St. Jacques, J.-M., Barrow, E. M., & Sauchyn, D. J. (2012). GCM projections for the Pacific Decadal Oscillation under greenhouse forcing for the early 21st century. *International Journal of Climatology*, *32*(9), 1423–1442.

Leitgeb, N. (2010). How safe is safe enough? Safety of Electromedical Devices. https://doi.org/10.1007/978-3-211-99683-6_2

Lempert, R. J. (2002). A new decision sciences for complex systems. *Proceedings of the National Academy of Sciences of the United States of America*, *99 Suppl 3*, 7309–7313.

Lempert, R. J., & Collins, M. T. (2007). Managing the risk of uncertain threshold responses: comparison of robust, optimum, and precautionary approaches. *Risk Analysis: An Official Publication of the Society for Risk Analysis*, *27*(4), 1009–1026.

Lempert, R. J., Groves, D. G., Popper, S. W., & Bankes, S. C. (2006). A General, Analytic Method for Generating Robust Strategies and Narrative Scenarios. *Management Science*, *52*(4), 514–528.

Lopez-Cantu, T., Prein, A. F., & Samaras, C. (2020). Uncertainties in Future U.S. Extreme Precipitation From Downscaled Climate Projections. *Geophysical Research Letters*, *47*(9), WO3410.

Lopez-Cantu, T., & Samaras, C. (2018). Temporal and spatial evaluation of stormwater engineering standards reveals risks and priorities across the United States. *Environmental Research Letters: ERL [Web Site]*, *13*(7), 074006.

Mallakpour, I., AghaKouchak, A., & Sadegh, M. (2019). Climate-Induced Changes in the Risk of Hydrological Failure of Major Dams in California. *Geophysical Research Letters*, *46*(4), 2130–2139.

Manocha, N., & Babovic, V. (2018). Sequencing Infrastructure Investments under Deep Uncertainty Using Real Options Analysis. In *Water* (Vol. 10, Issue 2, p. 229). https://doi.org/10.3390/w10020229

Mantua, N. J., Hare, S. R., Zhang, Y., Wallace, J. M., & Francis, R. C. (1997). A Pacific Interdecadal Climate Oscillation with Impacts on Salmon Production. In *Bulletin of the American Meteorological Society* (Vol. 78, Issue 6, pp. 1069–1079). https://doi.org/2.0.co;2">10.1175/1520-0477(1997)078<1069:apicow>2.0.co;2

Mays, L. W. (2010). *Water Resources Engineering*. John Wiley & Sons.

McGinnis, S., Nychka, D., & Mearns, L. O. (2015). A New Distribution Mapping Technique for Climate Model Bias Correction. In *Machine Learning and Data Mining Approaches to Climate Science* (pp. 91–99). https://doi.org/10.1007/978-3-319-17220-0_9

Mearns, L. O., McGinnis, S., Korytina, D., Arritt, R., Biner, S., Bukovsky, M., Chang, H. I., Christensen, O., Herzmann, D., Jiao, Y., & Others. (2017). The NA-CORDEX dataset, version 1.0. *NCAR Climate Data Gateway. Boulder (CO): The North American CORDEX Program*, *10*, D6SJ1JCH.

Melchers, R. E., & Beck, A. T. (2018). *Structural Reliability Analysis and Prediction*. John Wiley & Sons.

Perica, S., Kane, D., Dietz, S., Maitaria, K., Martin, D., Pavlovic, S., Roy, I., Stuefer, S., Tidwell, A., Trypaluk, C., Unruh, D., Yekta, M., Betts, E., Bonnin, G., Heim, S., Hiner, L., Lilly, E., Narayanan, J., Yan, F., & Zhao, T. (2012). Precipitation-frequency atlas of the United States. Volume 7 version 2.0. Alaska. https://repository.library.noaa.gov/view/noaa/22615

Presidential Document (2021). *Protecting Public Health and the Environment and Restoring Science To Tackle the Climate Crisis*. (2021, January 25). Presidential Documents. https://www.govinfo.gov/content/pkg/FR-2021-01-25/pdf/2021-01765.pdf

Ragno, E., AghaKouchak, A., Cheng, L., & Sadegh, M. (2019). A generalized framework for process-informed nonstationary extreme value analysis. *Advances in Water Resources*, *130*, 270–282.

Randall, F. A. (1976). The safety factor of structures in history. *Professional Safety*, 12–28.

Reynolds, R. W., Rayner, N. A., Smith, T. M., Stokes, D. C., & Wang, W. (2002). An Improved In Situ and Satellite SST Analysis for Climate. In *Journal of Climate* (Vol. 15, Issue 13, pp. 1609–1625). https://doi.org/2.0.co;2">10.1175/1520-0442(2002)015<1609:aiisas>2.0.co;2

Ropelewski, C. F., & Jones, P. D. (1987). An extension of the Tahiti–Darwin southern oscillation index. *Monthly Weather Review*, *115*(9), 2161–2165.

Sanders, B. F., & Grant, S. B. (2020). Re-envisioning stormwater infrastructure for ultrahazardous flooding. WIREs. Water, 7(2). https://doi.org/10.1002/wat2.1414

Salas, J. D., Obeysekera, J., & Vogel, R. M. (2018). Techniques for assessing water infrastructure for nonstationary extreme events: a review. *Hydrological Sciences Journal*, *63*(3), 325–352.

Schueler, T. R., & Claytor, R. A. (2000). Maryland stormwater design manual volumes I and II. *Center for Watershed Protection and Maryland Dep. of the Environment, Baltimore, Md*.

Smith, A. B. (2018). 2017 US billion-dollar weather and climate disasters: a historic year in context. *Report From Climate. Gov, Https://www. Climate. Gov/news-Features/blogs/beyond-data/2017-Us-Billion-Dollar-Weather-and-Climate-Disasters-Historic-Year, Accessed June*, *28*, 2019.

Spiegelhalter, D. J., Best, N. G., Carlin, B. P., & van der Linde, A. (2002). Bayesian measures of model complexity and fit. In *Journal of the Royal Statistical Society: Series B (Statistical Methodology)* (Vol.

64, Issue 4, pp. 583–639). https://doi.org/10.1111/1467-9868.00353

Stehfest, E., van Zeist, W.-J., Valin, H., Havlik, P., Popp, A., Kyle, P., Tabeau, A., Mason-D'Croz, D., Hasegawa, T., Bodirsky, B. L., Calvin, K., Doelman, J. C., Fujimori, S., Humpenöder, F., Lotze-Campen, H., van Meijl, H., & Wiebe, K. (2019). Key determinants of global land-use projections. *Nature Communications*, *10*(1), 2166.

Taylor, K. E., Stouffer, R. J., & Meehl, G. A. (2012). An Overview of CMIP5 and the Experiment Design. *Bulletin of the American Meteorological Society*, *93*(4), 485–498.

Trak, B. (2017). How safe is the "factor of safety" concept in geotechnical practice? In Geo-Risk 2017. Reston, VA: American Society of Civil Engineers. https://doi.org/10.1061/9780784480724.028

Tran Huu D. (2016). Markov-Based Reliability Assessment for Hydraulic Design of Concrete Stormwater Pipes. *Journal of Hydraulic Engineering*, *142*(7), 06016005.

Underwood, B. S., Mascaro, G., Chester, M. V., Fraser, A., Lopez-Cantu, T., & Samaras, C. (2020). Past and present design practices and uncertainty in climate projections are challenges for designing infrastructure to future conditions. Journal of Infrastructure Systems, 26(3), 04020026.

Vahedifard, F., AghaKouchak, A., Ragno, E., Shahrokhabadi, S., & Mallakpour, I. (2017). Lessons from the Oroville dam. In *Science* (Vol. 355, Issue 6330, pp. 1139.2–1140). https://doi.org/10.1126/science.aan0171

Viterbo, F., Mahoney, K., Read, L., Salas, F., Bates, B., Elliott, J., Cosgrove, B., Dugger, A., Gochis, D., & Cifelli, R. (2020). A Multiscale, Hydrometeorological Forecast Evaluation of National Water Model Forecasts of the May 2018 Ellicott City, Maryland, Flood. *Journal of Hydrometeorology*, *21*(3), 475–499.

Winsemius, H. C., Jeroen C J, van Beek, L. P. H., Bierkens, M. F. P., Bouwman, A., Jongman, B., Kwadijk, J. C. J., Ligtvoet, W., Lucas, P. L., van Vuuren, D. P., & Ward, P. J. (2015). Global drivers of future river flood risk. *Nature Climate Change*, *6*(4), 381–385.

Wise, E. K., Wrzesien, M. L., Dannenberg, M. P., & McGinnis, D. L. (2015). Cool-Season Precipitation Patterns Associated with Teleconnection Interactions in the United States. In *Journal of Applied*

*Meteorology and Climatology* (Vol. 54, Issue 2, pp. 494–505). https://doi.org/10.1175/jamc-d-14-0040.1

Woldemeskel, F. M., Sharma, A., Sivakumar, B., & Mehrotra, R. (2016). Quantification of precipitation and temperature uncertainties simulated by CMIP3 and CMIP5 models. *Journal of Geophysical Research, D: Atmospheres*, *121*(1), 3–17.

Wong, T. E., & Keller, K. (2017). Deep Uncertainty Surrounding Coastal Flood Risk Projections: A Case Study for New Orleans. In *Earth's Future* (Vol. 5, Issue 10, pp. 1015–1026). https://doi.org/10.1002/2017ef000607

Wong, T. E., Klufas, A., Srikrishnan, V., & Keller, K. (2018). Neglecting model structural uncertainty underestimates upper tails of flood hazard. In *Environmental Research Letters* (Vol. 13, Issue 7, p. 074019). https://doi.org/10.1088/1748-9326/aacb3d

Wright, D. B., Bosma, C. D., & Lopez-Cantu, T. (2019). US hydrologic design standards insufficient due to large increases in frequency of rainfall extremes. *Geophysical Research Letters*, *46*(14), 8144–8153.

Zarekarizi, M., Srikrishnan, V., & Keller, K. (2020). Neglecting uncertainties biases house-elevation decisions to manage riverine flood risks. *Nature Communications*, *11*(1), 5361.

**Figures:**

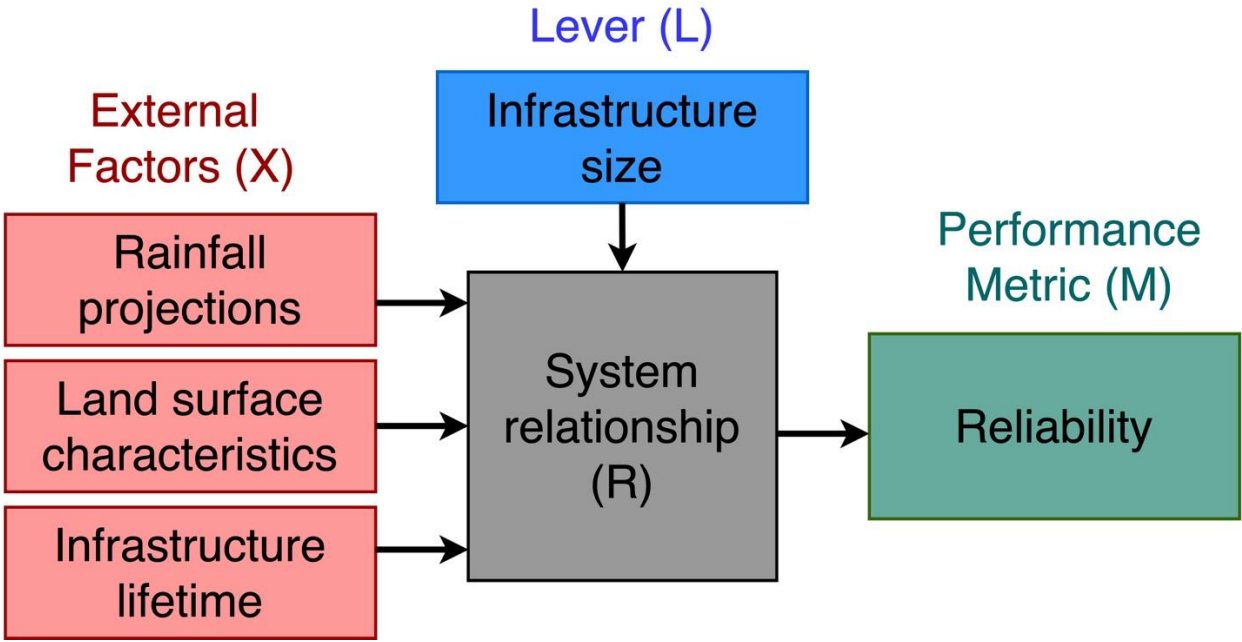

**Figure 1**: **Flow diagram for the decision-analysis.** The XLRM framework (Lempert et al., 2006) illustrates the relationship between the exogenous external factors (X), lever (L) (i.e., infrastructure size), the system relationship (R) and the performance metric (M).

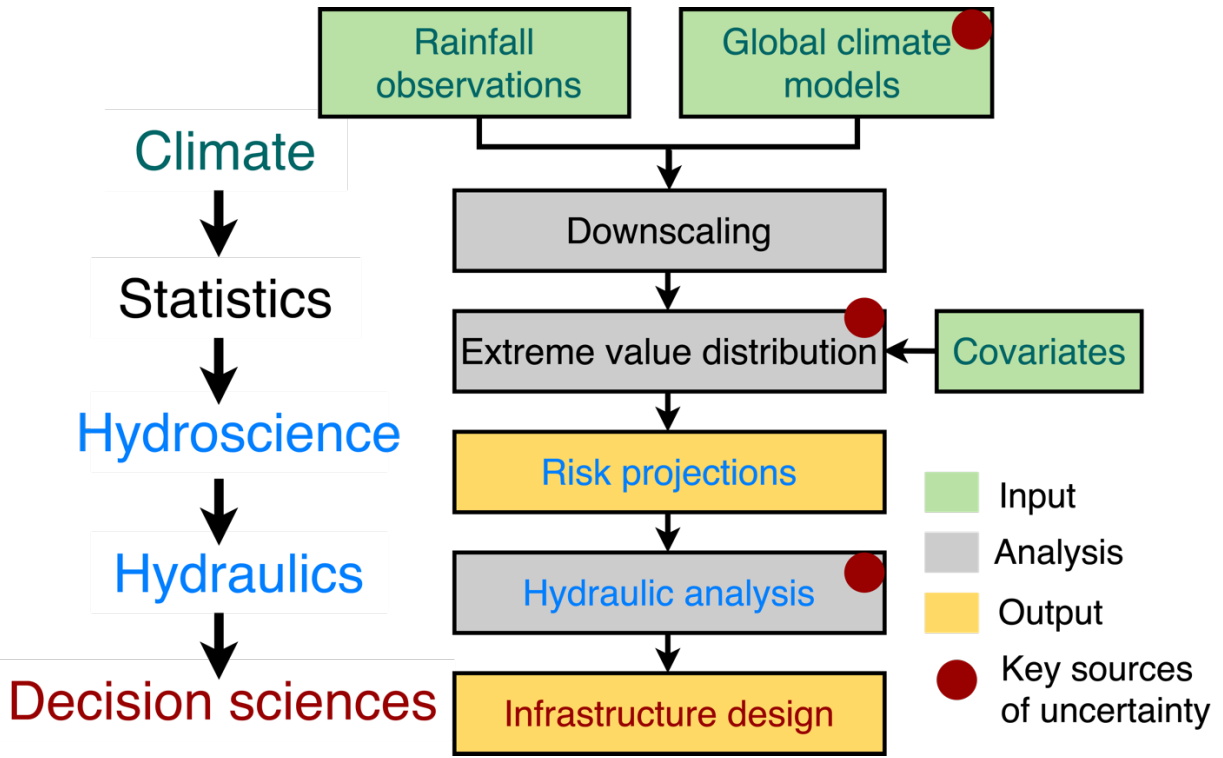

**Figure 2**: Diagrammatic representation of the design analysis for stormwater drainage size.

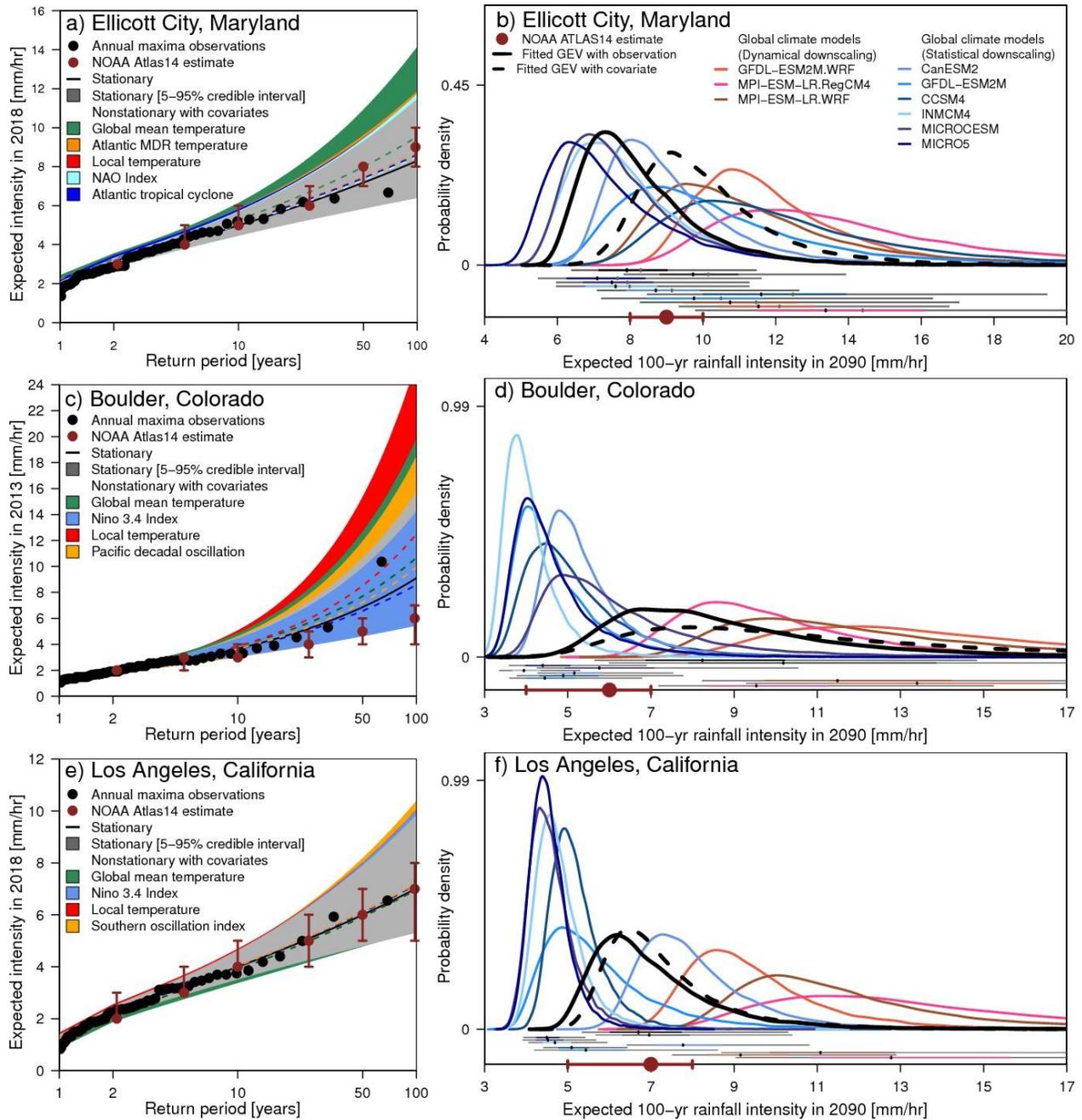

**Figure 3: Expected current and future rainfall return periods. a,** Expected rainfall intensity in Ellicott City, Maryland in 2018 for different return periods under the stationary and nonstationary assumptions. The dotted lines are the expected return levels and the corresponding color bound shows the 5%-95% credible interval. The brown dots represent the United States National Oceanic and Atmospheric Administration (NOAA) Atlas 14 estimates (Perica et al., 2012); and the corresponding error bars represent the 90% confidence bound. We compute nonstationary rainfall

intensity using Generalized Extreme Value (GEV) distribution fitted with different covariates. Selected covariates are global mean temperature, Atlantic Main Development Region (MDR) temperature, local temperature, North Atlantic Oscillation (NAO) index and number of Atlantic tropical cyclones. We calculate the Akaike Information Criterion (AIC) and Deviance Information Criterion (DIC) to assess the goodness-of-fit for GEV distribution. For the observed rainfall intensity, the selected metrics choose the GEV distribution with MDR temperature as the best nonstationary model. **b,** Projected daily rainfall intensity in 2090 using GEV distribution fitted with historical observation (1951-2018) and climate model outputs (2020-2090) using MDR temperature as a covariate. We use the North American Coordinated Regional Downscaling Experiment (NA-CORDEX)( Mearns et al., 2017) project dataset, which are dynamically downscaled products. The Multivariate Adaptive Constructed Analogs (MACA) (Abatzoglou & Brown, 2012) data sets are statistically downscaled subset of the coupled model intercomparison project phase 5 (CMIP5) Global climate models (GCMs). **c,** Expected rainfall intensity in Boulder, Colorado in 2013 for different return periods under the stationary and nonstationary assumptions. Selected covariates are local temperature than the global mean surface temperature, the Pacific decadal oscillation index and the Nino 3.4 index. For the observed rainfall intensity, the selected metrics choose the GEV distribution with local temperature as the best nonstationary model. **d,** Projected daily rainfall intensity in 2090 using GEV distribution fitted with historical observation (1951-2018) and climate model outputs (2020-2090) using local temperature as a covariate. **e,** Expected rainfall intensity in Los Angeles, California in 2018 for different return periods under the stationary and nonstationary assumptions. Selected covariates are global mean surface temperature, local temperature, the Nino 3.4 index and the Southern oscillation index. We choose the Nino 3.4 index as a covariate to project extreme rainfall intensity in Los Angeles. **f,** Projected daily rainfall intensity in 2090 using GEV distribution fitted with historical observation (1951-2018) and climate model outputs (2020-2090) using the Nino 3.4 index as a covariate.

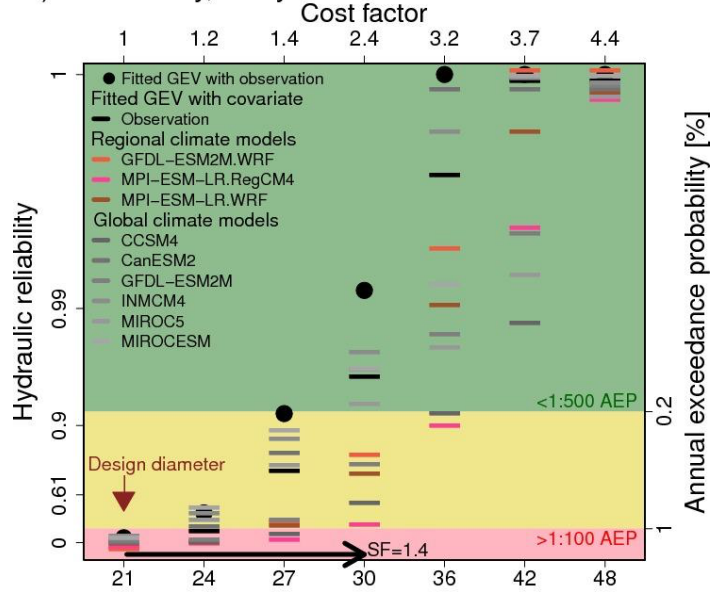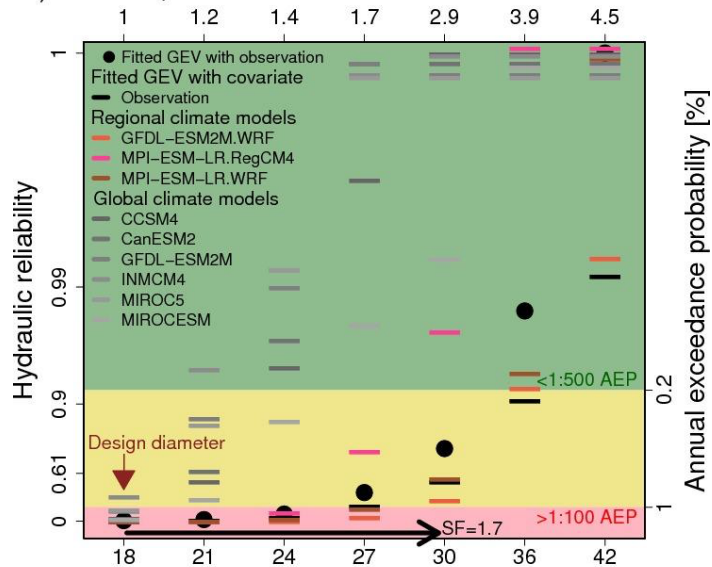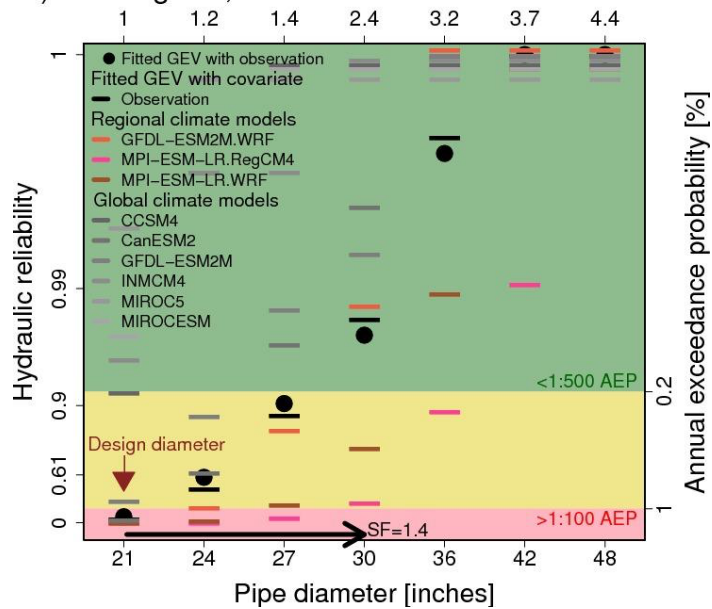

**Figure 4: Lifetime hydraulic reliability of stormwater drainage pipe.** Lifetime hydraulic reliability of stormwater pipe conditioned on the deeply uncertain extreme rainfall projections, surface imperviousness and pipe lifetime. We assume a concrete pipe with Manning's roughness, $n$=0.013, and a contributing drainage area of 0.25 km$^2$. Cost factors (Heaney et al., 2002) (top x-axis) are relative to the design diameter. Design diameter is the resulting diameter with daily rainfall intensity from the United States National Oceanic and Atmospheric Administration (NOAA) Atlas 14 estimates (Perica et al., 2012) for the selected stations: a) Baltimore/Washington International Thurgood Marshall Airport, Maryland, b) Boulder Municipal Airport, Colorado, and c) Los Angeles International Airport, California. Safety factor (SF) (Randall, 1976) is a multiplier in a stormwater pipe diameter obtained from the standard design guidance. Safety factor allows to achieve the intended hydraulic reliability that is robust to the considered deep uncertainty.